\documentclass[journal]{IEEEtran}
\usepackage{graphics}
\usepackage{epsfig,subfigure}
\usepackage{amsmath,amssymb,mathrsfs,txfonts}

\begin{document}
\title{An OFDM Signal Identification Method for Wireless Communications Systems}

\author{\IEEEauthorblockN{Ali Gorcin\IEEEauthorrefmark{1},
Huseyin Arslan\IEEEauthorrefmark{1}\IEEEauthorrefmark{2},\\
\IEEEauthorblockA{\IEEEauthorrefmark{1}Department of Electrical Engineering,
University of South Florida, 4202 E. Fowler Ave., ENB-118, Tampa, FL, 33620, USA}\\
\IEEEauthorblockA{\IEEEauthorrefmark{2} School of Engineering and Natural Sciences, Istanbul Medipol University, Beykoz, Istanbul, 34810, Turkey}\\
Emails: agorcin@mail.usf.edu, arslan@usf.edu}}
\maketitle
\footnote{Accepted for publication in the IEEE Transactions on Vehicular Technology. Copyright (c) 2013 IEEE. Personal use of this material is permitted. However, permission to use this material for any other purposes must be obtained from the IEEE by sending a request to pubs-permissions@ieee.org.}
\begin{abstract}
Distinction of OFDM signals from single carrier signals is highly important for adaptive receiver algorithms and signal identification applications. OFDM signals exhibit Gaussian characteristics in time domain and fourth order cumulants of Gaussian distributed signals vanish in contrary to the cumulants of other signals. Thus fourth order cumulants can be utilized for OFDM signal identification. In this paper, first, formulations of the estimates of the fourth order cumulants for OFDM signals are provided. Then it is shown these estimates are affected significantly from the wireless channel impairments, frequency offset, phase offset and sampling mismatch. To overcome these problems, a general chi-square constant false alarm rate Gaussianity test which employs estimates of cumulants and their covariances is adapted to the specific case of wireless OFDM signals. Estimation of the covariance matrix of the fourth order cumulants are greatly simplified peculiar to the OFDM signals. A measurement setup is developed to analyze the performance of the identification method and for comparison purposes. A parametric measurement analysis is provided depending on modulation order, signal to noise ratio, number of symbols, and degree of freedom of the underlying test. The proposed method outperforms statistical tests which are based on fixed thresholds or empirical values, while a priori information requirement and complexity of the proposed method are lower than the coherent identification techniques.
\end{abstract}

\section{Introduction}
Modulation identification has always been a part of communications surveillance systems and electronic warfare, however gained more importance with the adaptive communication systems. Seamless and efficient communications should be achieved between a wide variety of communications systems which may employ different modulation techniques. Therefore, simple, robust, and fast modulation identification methods are required at the implementation of adaptive and intelligent receivers. Blind receiver design algorithms and signal transmission procedures can also benefit from the modulation recognition information. 

Improving the efficiency of the wireless communications systems, on the other hand, has substantial importance due to the increasing demand for higher capacity. Ascending number of users, networks, and deployment of a wide variety of systems led the wireless communications towards heterogeneous networks, small cell deployments and implementation of dynamic spectrum access techniques. Therefore emerging technologies require as much information as possible about the communications medium. Wireless communications measurement systems should also adapt themselves to the new wireless communications paradigm by providing more information such as modulation type, carrier number, bandwidth, and other waveform parameters automatically. To this end, radiocommunication sector of International Telecommunication Union (ITU-R) initialized an incentive, and published the ``Technical identification of digital signals'' document as the main guideline for the wireless communications measurement systems to identify unexpected or unlicensed emissions, to detect interference sources, and to provide waveform parameters via signal identification \cite{SM.2152}.

\subsection{Related Work}
When the contemporary communications systems are considered, orthogonal frequency-division multiplexing (OFDM) is a multi-carrier (MC) multiplexing scheme which maintains adaptive communications features by employing sub-carriers in a flexible way. Deployments of OFDM based systems are increasing rapidly. Distinction of OFDM signals from the single carrier (SC) signals is an important issue from the adaptive transceiver design, signal identification and spectrum monitoring perspectives. Initial research on this field indicates that modulation identification or recognition methods for OFDM systems can be classified into two groups; coherent techniques which include maximum likelihood (ML) modulation classifiers along with other methods and non-coherent statistical methods which do not require a priori information, in general. Coherent techniques are based on the assumption that certain characteristics of the OFDM signals are known to the receiver. Statistical methods mostly depend on high order statistics (HOS) of cumulant or moment estimates of the received signals. The background of cumulant based modulation classification laid down by \cite{Swami} for fourth and sixth order cumulants. Later, fourth order cumulants are adapted to OFDM modulation identification \cite{Grimaldi}-\nocite{Shi}\cite{Liu}. Moreover, fourth order moments of OFDM signals are combined with the amplitude estimation to improve signal detection in \cite{Ulovec}. On the other hand, support vector machines in combination with specific fourth order cumulants are used for the same purpose in \cite{HWang}. Tabular values of high order moments which were implemented for SC modulation identification in \cite{WeiDai} are adapted to a minimum mean-square error (MMSE) estimator for OFDM signals in \cite{Wang}. A low complexity single carrier modulation identification algorithm based on fourth order cumulants for multiple-input and multiple-output systems is also proposed in \cite{Muhlhaus}.

The ML modulation classification algorithms are known with better estimation results when compared to statistical methods, however they suffer from computational complexity and require channel estimation \cite{Leinonen}. On the other hand, bootstrap technique is introduced as a pattern recognition based method in \cite{FWang}. This method conducts SC modulation separation along with the SC/MC distinction. A wavelet transform algorithm is proposed to extract transient characteristics of OFDM signals in \cite{Zhang}. Moreover, time-frequency representation of the wireless signals is employed to distinguish OFDM signals from SC frequency-shift keying (FSK) and phase-shift keying (PSK) signals in \cite{Haq}. OFDM signals have unique second order cyclostationary features and a feature extraction algorithm based on the magnitude of the cyclic cumulants of OFDM signals is implemented in \cite{Punchihewa}. Furthermore, second order cyclostationary  features of block transmitted-single carrier linearly digitally (BT-SCLD) modulated signals are utilized for distinction from OFDM signals and for parameter estimation in \cite{QZhang}. Another second order cyclostationary  algorithm to detect and classify both worldwide interoperability for microwave access (WiMAX) and long term evolution (LTE) signals is developed in the context of cognitive radio systems and spectrum awareness concept in \cite{Ala}. An analysis and comparison of second order cyclostationary, matched filter, MMSE, and normalized kurtosis methods can be found in \cite{Bouzegzi} and for a general overview of coherent techniques, \cite{Hazza} can be consulted. 

OFDM signals also exhibit time-domain Gaussianity. This is due to fact that the assignment of random data over orthogonal sub-carriers in a simultaneous way can be assumed as a composition of large number of independent, identically distributed (i.i.d.) random variables and central limit theorem implies Gaussian distribution for large enough data sets. A Gaussianity test based on empirical distribution function is introduced for OFDM systems in \cite{Li}. In addition to that, a general chi-squared constant false alarm rate (CFAR) Gaussianity test based on the estimates of the third and fourth order cumulants \cite{Giannakis} is adapted to OFDM signals without considering the channel effects in \cite{Akmouche}.

\subsection{Proposed Method}
When the coherent methods are taken into account, there are two main drawbacks when compared to the statistical methods: first, the dependence on the specific signal parameters such as symbol duration, cyclic-prefix size, active carrier numbers for the identification of each signal type and standard. Second, the algorithm complexity can increase the software and/or hardware complexity thus complicate the feasibility of implementation of the method. Moreover, complexity issues can lead to extension of identification duration. On the other hand, it is known that statistical methods are simpler than the coherent methods however they are highly susceptible to the changes in the wireless channel conditions along with other impairments such as frequency offset, phase offset due to sampling mismatch and other factors \cite{Gorcin}.  Moreover, the statistical tests in the literature depend on either fixed thresholds or empirical values of various HOS. However HOS can significantly fluctuate under wireless channel conditions and the identification performances of these tests degrade significantly.

It is known that considering all combinations of possible lags is a computationally consuming process for cumulant based statistical classification methods. However peculiar to the OFDM signals, inclusion of certain lags in the analysis is sufficient to retain Gaussianity. Therefore in this study, first the fourth order cumulant equations for the OFDM signals are provided for limited number and combination of lags and lag distances. Second, in \cite{Gorcin} a general analysis of the effects of wireless impairments over the fourth order cumulants of OFDM signals were conducted. Furthermore in this paper, the fourth order cumulant estimates including the wireless effects are provided. These equations imply fundamental impact of impairments over the cumulant values. Third, an OFDM signal identification method which is based on a modified chi-squared Gaussianity test is proposed to distinguish MC OFDM signals from SC signals. The method employs the estimates of the fourth order cumulants and their covariance matrix in the decision process. Therefore, a certain level of protection is achieved against the effects of multipath fading channels, when compared to the statistical methods which are based on only the estimates of cumulants or moments of wireless signals. Moreover as detailed in Section \ref{FOCoOS}, instead of fixed numerical values, the proposed identification method utilizes the degree of freedom of the underlying chi-squared test which provides relative isolation from the effects of wireless impairments. 

Fourth, the proposed method includes the computation of the estimate of the covariance matrix of fourth order cumulants. In general, this is a complex and detailed operation because the combinations of products of the second and fourth order moments are included in the estimation process. However, in this paper the estimation process is reduced down to three particular equations peculiar to the OFDM signals. The only parameter required is the symbol duration which determine the degree of freedom of the modified chi-squared test. The numerical values that satisfy the confidence levels for correct detection can be found iteratively thus the test can become fully blind. Fifth, a measurement setup is developed and the proposed method is implemented to achieve parametric performance analysis in a realistic way. A signal generation software is developed and integrated to the signal generator to enable parametric flexibility during the signal transmission process. Such an approach led to detailed parametric analysis of the proposed method based on modulation order, signal to noise ratio (SNR), number of symbols, and degree of freedom of the underlying test against FSK, PSK, and quadrature amplitude modulation (QAM) SC signals. The detection performance of the proposed method is also compared with three statistical methods which are based on HOS and a coherent identification method which employs second order cyclostationary features of OFDM signals for identification.

The OFDM signal model which is used at the implementation of the proposed method is defined in Section \ref{SM}. The fourth order cumulant expressions for OFDM signals, discussion on the effects of wireless impairments on the cumulants, the modified chi-squared test which is employed by the proposed method, the decision mechanism, and the flow of the identification method are provided in Section \ref{FOCoOS}. The estimates of the covariances of the cumulants are derived in the Appendix. The measurement setup is detailed in Section \ref{MS}. Parametric performance analysis of the proposed signal identification method based on the measurements is given in Section \ref{MR}. Finally, conclusion constituted the Section \ref{C}.

\section{Signal Model}
\label{SM}
Inverse discrete Fourier transform (IDFT) is employed by OFDM systems for transmission. The baseband continuous-time OFDM signal is given by
\begin{equation}\label{eqn:1}
    x_n(t) = \sum^K_{k=1} S_n(k)e^{j\frac{2\pi\alpha(k)t}{T_D}}\qquad-T_G\leq t < T_D, 
\end{equation}
\noindent where $j$ at the exponential term is the imaginary unit, $\alpha$ is the set of K subcarriers in total, $S_n(k)$ is the $n^{th}$ OFDM data symbol which is transmitted at the $k^{th}$ subcarrier from the set of $\alpha(k)$, $T_D$ is the useful data duration, $T_G$ is the cyclic prefix (CP) duration, and the total OFDM symbol duration is  $T_S = T_D + T_G$. The transmitted total signal can be written as
\begin{equation}\label{eqn:2}
    x(t) = \sum^\infty _{n = -\infty} x_n(t - nT_S).
\end{equation}

The signal is modulated after digital analog (D/A) conversion and passed through mobile radio channel. The channel can be modeled as a time-variant linear filter
\begin{equation}\label{eqn:3}
    h(t) = \sum_{m = 1}^{L}h_m(t)\delta(t -\tau_m),
\end{equation}
\noindent where $L$ is the number of taps and $\tau_i$ is the excess delay for each tap. It is assumed that the taps are sample spaced and the channel is constant for a symbol but time-varying across multiple OFDM symbols. Therefore, baseband model of the received signal after down conversion also considering the channel model given in eqn. \eqref{eqn:3} can be given by
\begin{equation}\label{eqn:5}
    y(t) = e^{j\theta_t}e^{j2\pi\xi t}\sum_{m = 1}^{L} x(t - \tau_m)h_m(t) + n(t),
\end{equation}
\noindent where $\xi$ is the carrier frequency offset due to inaccurate frequency synchronization, $\theta_t$ is the phase offset, and $n(t)$ corresponds to additive white Gaussian noise (AWGN) sample with zero mean and variance of $\sigma_n^2$. The received signal is sampled with sampling time of $\Delta t$ at the analog to digital converter (ADC) and discrete-time received signal can be represented by
\begin{equation}\label{eqn:6}
    y(i) = e^{j\Delta\theta_i}e^{j2\pi\Delta fi}\sum_{m = 1}^{L} x(i\Delta t - \tau_m)h_m(i\Delta t) + n(i),
\end{equation}
\noindent where $i = 1,\ldots,N$, $\Delta\theta_i$ and $\Delta f$ are normalized phase and carrier frequency offsets. Models of FSK, PSK, and QAM modulated SC signals are not listed here, but can be found in \cite{Grimaldi} and references therein. Finally, at the initial stage of the identification process, it is assumed that the receiver is focused on a single channel which may be utilized by a SC or a MC OFDM signal comprised of sub-carriers, the bandwidth and carrier frequency are estimated, the signal is filtered, down-converted and sampled beforehand. It is also assumed that in case of signals that are overlapping on wireless spectrum, before the initialization of the identification process, the proposed method can benefit from the signal separation processes introduced in \cite{Gorcin2}.

\section{Proposed Identification Method}
\label{FOCoOS}
The proposed method benefits from the high order cumulants of the received signals for signal identification. The cumulants of a process are defined as the generalization of the autocorrelation function $E\{y(i)y(i+i_1)\}$. Thus, the general formulation of third and fourth order cumulants for the stationary random processes with zero mean is given by
\begin{equation}\label{eqn:7}
    c_{\mathit{3}y}(i_1,i_2) \triangleq \mathrm{E}\{y(i)y(i+i_1)y(i+i_2)\}
\end{equation}
\noindent and
\begin{align}\label{eqn:8}
     c_{\mathit{4}y}(i_1,i_2,i_3) \triangleq &\ \mathrm{E}\{y(i)y(i+i_1)y(i+i_2)y(i+i_3)\}\nonumber\\
                                    & - c_{\mathit{2}y}(i_1)c_{\mathit{2}y}(i_2 - i_3) - c_{\mathit{2}y}(i_2)c_{\mathit{2}y}(i_3 - i_1)\nonumber\\
                                    & - c_{\mathit{2}y}(i_3)c_{\mathit{2}y}(i_2 - i_1),
\end{align}
\noindent where $c_{2y}(i_1) \triangleq E\{y(i)y(i+i_1)\}$. The third and fourth order cumulants have different characteristics; the \textit{k}th order cumulants will vanish for $\mathit{k}>3$ if $y(i)$ is Gaussian \cite{Brillinger2}. However, third order cumulants can converge to zero although $y(i)$ is non-Gaussian but symmetrically distributed \cite{Mendel}. Therefore, third order cumulants will be ignored and fourth order cumulants will be analyzed for OFDM signal identification. On the other hand, when the calculation of the estimates of the cumulants are considered, the estimated cumulants, $\hat{c}_{ky}$, should be absolutely summable \cite{Brillinger1} and the signal model for wireless OFDM signals which is given in eqn. \eqref{eqn:6} satisfies this condition. Under these conditions, the estimate autocorrelation becomes $\hat{c}_{2y}(i_1) = \frac{1}{N} \sum_{i = 1}^{N-i_1}y(i)y(i+i_1)$, and 4th-order cumulant estimate is given by
\begin{align}\label{eqn:12}
     \hat{c}_{\mathit{4}y}(i_1,i_2,i_3) \triangleq &\ \frac{1}{N} \sum_{i = 1}^{N-i_1}y(i)y(i+i_1)y(i+i_2)y(i+i_3)\nonumber\\
                                    & - \hat{c}_{\mathit{2}y}(i_1)\hat{c}_{\mathit{2}y}(i_2 - i_3) - \hat{c}_{\mathit{2}y}(i_2)\hat{c}_{\mathit{2}y}(i_3 - i_1)\nonumber\\
                                    & - \hat{c}_{\mathit{2}y}(i_3)\hat{c}_{\mathit{2}y}(i_2 - i_1), \quad (i_1,i_2,i_3) \in I_4^N.
\end{align}

The region of all combinations of all lags is given by $I^{\infty}_\mathit{k} \triangleq \{0\leq i_{k-1}\leq\dots\leq i_1\leq\infty\}$. However, it is shown in \cite{Giannakis2} that cumulant lags that comprise distribution characterization of the series are finite and it's not required to estimate the cumulant values for the majority of the possible lag combinations. Therefore, the region of the lags that should be taken into consideration for the analysis is given by
\begin{equation}\label{eqn:11}
    I_4^N = \{0\leq i_3\leq i_2\leq i_1\leq N\},
\end{equation}
\noindent and the sample estimates in eqn. \eqref{eqn:12} should only be calculated over $I_4^N$ as defined in eqn. \eqref{eqn:11}. The Gaussianity test will use lags of $\hat{c}_{\mathit{4}y}$ that can be collected into an $N_c\times 1$ vector. This vector constitutes a 3 dimensional triangular region and its length is given by $N_c = N(N+1)(N+2)/6$.
\subsection{Fourth Order Cumulants for OFDM Signals}
\label{FOCOS}
When the wireless signals are considered, the length of the data can be orders of thousands or tens of thousands of samples depending on the sampling rate and the recording time. Defining the $I_4^N$ lag region and computing the components of vector $\hat{c}_{\mathit{4}y}$ can become impractical in many cases. 

The data set can be divided into smaller sections and processed in parallel or the data set can be shortened. However, even though the choice of $N$ and consequently $N_c$ is application dependent, in practice when a strong non-Gaussianity (\textit{e.g.}, due to the multipath fading channel) present in short records \textit{i.e}, $N < 200$, this will weaken the Gaussianity assumption and will lead to unreliability for the signal identification methods. Moreover, even for the shortest data sets, the size of the $\hat{c}_{\mathit{4}y}$ vector can become very large and the computation of the cumulants can be impractical. However, \cite{Martret} and \cite{Akmouche} showed that, first, shrinking the lag region into $I_4^M = \{i_3=0\leq i_2 = i_1 \leq M\}$ does not lead to loss of significant distribution information for communications signals. The terms that are left over have very small influence on the features of the modulation type of the signal when compared to the set in the hand. Second, the lag distance can also be limited with $M \approx 1.5 T_s$ where $T_s$ is the symbol duration for the signal under test. Therefore the cumulants should be evaluated only for the set of lags of $(\lambda,\lambda,0)$ where $i_1 = i_2 =\lambda \in [0,1.5T_s]$. Under these conditions eqn. \eqref{eqn:12} can be written as
\begin{align}\label{eqn:15}
     \hat{c}_{\mathit{4}y}(\lambda,\lambda,0) \triangleq &\ \frac{1}{N} \sum_{i = 1}^{N-\lambda}y(i)y(i+\lambda)y(i+\lambda)y(i)\nonumber\\
                                    & - \hat{c}_{\mathit{2}y}(\lambda)\hat{c}_{\mathit{2}y}(\lambda) - \hat{c}_{\mathit{2}y}(\lambda)\hat{c}_{\mathit{2}y}(-\lambda)\nonumber\\
                                    & - \hat{c}_{\mathit{2}y}(0)\hat{c}_{\mathit{2}y}(0), \quad (i_1,i_2,i_3) \in I_4^M,
\end{align}
\noindent and the open form becomes
\begin{align}
\hat{c}_{\mathit{4}y}(\lambda,\lambda,0) =\nonumber\\
\ \frac{1}{N} \sum_{i = 1}^{N-\lambda}\ & y^{2}(i)y^{2}(i+\lambda) - \Bigg[\frac{1}{N} \sum_{i = 1}^{N-\lambda}y(i)y(i+\lambda)\Bigg]^2\tag{11.1}\label{eqn:394}\\
                                     - \Bigg[\frac{1}{N} \sum_{i = 1}^{N-\lambda}\ & y(i)y(i+\lambda)\Bigg]\Bigg[\frac{1}{N} \sum_{i = \lambda+1}^{N}y(i)y(i-\lambda)\Bigg]\tag{11.2}\\\label{eqn:3943}
                                      - \Bigg[\frac{1}{N} \sum_{i = 1}^{N}\ & y^2(i)\Bigg]^2.\tag{11.3}\\\label{eqn:39443}
\end{align}

When the signal model in the eqn.~\eqref{eqn:6} is replaced into the eqn.~\eqref{eqn:394}, the statements that are given in the eqn.~\eqref{eqn_dbl_x} is achieved. Instead of expanding whole eqn.~\eqref{eqn:6}, eqn.~\eqref{eqn_dbl_x} is sufficient for the analysis at the next step. The results can then be generalized for the rest of the eqn.~\eqref{eqn:6}. Note that the noise component $n(i)$ is independent from the rest of the terms that constitute the received signal and vanishes due to its Gaussianity. Therefore any term that included noise component is eliminated from the expressions given in eqn. \eqref{eqn_dbl_x}. When further mathematical operations on eqn. \eqref{eqn_dbl_x} is sought, the HOS theory \cite{Mendel} indicates that if $\eta_u$, where $u = 1,\ldots,k-1$, are constants then
\begin{equation}\label{eqn:111}
    c_{\mathit{k}y}(\eta_1i_1,\ldots,\eta_{k-1}i_{k-1}) = \Bigg(\prod_{u=1}^{k-1}\eta_u\Bigg)c_{\mathit{k}y}(i_1,\ldots,i_{k-1}),
\end{equation}
\noindent and based on this property of cumulants, for the linear and time-invariant systems it can be written that
\begin{equation}\label{eqn:9}
    y(i) = \sum_{j = -\infty}^{\infty} x(i - j)h(j) + n(i), \qquad \sum_{i = -\infty}^{\infty} |h(i)| < \infty
\end{equation}
\noindent and \textit{k}-th-order cumulants becomes \cite{Brillinger1}
\begin{align}\label{eqn:10}
    \hat{c}_{\mathit{k}y}(i_1,i_2,i_3,\dots,i_{k-1}) =\nonumber\\
     \hat{c}_{\mathit{k}x}(0,\ldots,0)\sum_{i = -\infty}^{\infty}& h(i)h(i+i_1)\dots\ h(i+i_{k-1}).
\end{align}

Assuming that the $x(i)$ is i.i.d., in case of non-Gaussian signals, zero-lag fourth order cumulants will converge to the finite moments of $\hat{c}_{4x}(0,0,0) \neq 0$ and in case of Gaussian distributed signals $\hat{c}_{4x}(0,0,0) = 0$. Thus $\hat{c}_{\mathit{4}y}(0,0,0)$ will also converge to zero theoretically. However, this is not the case for wireless signals because, the signal model given in Section \ref{SM} implies that wireless channel is time-variant across multiple symbols. Moreover, phase and frequency offset can not be assumed constant over the period of $N$ symbols which can extend to a couple of hundreds as indicated in Section \ref{FOCOS}. Therefore it is not possible to shift the channel coefficients, phase and frequency offset components out of the cumulant equations. The i.i.d. property will not perfectly hold because of these additional components introduced to each sample of the signal when compared to eqn.(\ref{eqn:1}). Hence the high order cumulants of wireless signals will not only be determined by the transmitted signal, but also will depend on the communications medium characteristics. Furthermore, the probabilistic convergence of estimated cumulants $\hat{c}_{\mathit{k}y}$ to the $c_{\mathit{k}y}$ can be possible when $y(i)$ samples are independent and well separated in time. This is called mixing conditions. However, the autocorrelation $\hat{c}_{\mathit{2}y}(i_1)$ is estimated using sample averaging because instead of the series $y(t)$, the samples of the original series $y(i)$ are employed in the process. Moreover, there is no synchronization process involved in the sampling stage. Therefore an irreducible error is introduced to the estimation of the cumulants. Under these conditions $\hat{c}_{\mathit{4}y}$ will not converge perfectly to zero and can take different numerical values based on the channel conditions and communications medium. On the other hand, when the fourth order cumulants are written in the form of
\begin{align}\label{eqn:989}
    \hat{c}_{\mathit{4}y}(i_1,i_2,i_3) = \mathrm{E}&\{y(i)y(i+i_1)y(i+i_2)y(i+i_3)\}\nonumber\\ &- \mathrm{E}\{g(i)g(i+i_1)g(i+i_2)g(i+i_3)\},
\end{align}
\noindent where $\{g(i)\}$ is a Gaussian random process, it can be inferred from the eqn. \eqref{eqn:989} that fourth order cumulants provide a measure of distance of the distribution of the signal samples form Gaussianity. Considering the fact that the transmitted SC signals will also be affected from same propagation characteristics and consequently their estimates of HOS will change, it is still plausible to assume that estimates of the fourth order cumulants of OFDM signals will converge to \textit{closer} numerical values to zero when compared to SC carrier signals. However, under wireless channel conditions and propagation characteristics, the statistical identification methods which are based on fixed thresholds or empirical numerical values such as \cite{Shi,Ulovec,WeiDai,Wang} will fail to identify OFDM signals consistently. To this end, a new approach is required to the identification of OFDM signals for wireless communications systems.
\begin{figure*}[!t]
\normalsize
\begin{align}\label{eqn_dbl_x}
\frac{1}{N}\sum_{i = 1}^{N-\lambda}\!y^{2}(i)y^{2}(i+\lambda)&- \Bigg[\frac{1}{N}\sum_{i = 1}^{N-\lambda}\!y(i)y(i+\lambda)\Bigg]^2\!=\frac{1}{N}\sum_{i = 1}^{N-\lambda}\!\!e^{2j(\Delta\theta_i+\Delta\theta_{i+\lambda})}e^{j2\pi\Delta fi}\Bigg(\sum_{m = 1}^{L}\!\!x(i\Delta t - \tau_m)h_m(i\Delta t)\Bigg)^2\Bigg(\sum_{m = 1}^{L}\!\!x((i+\lambda)\Delta t - \tau_m) h_m((i+\lambda)\Delta t)\Bigg)^2\nonumber\\&-  \Bigg[\frac{1}{N}\sum_{i = 1}^{N-\lambda}\!e^{2j(\Delta\theta_i+\Delta\theta_{i+\lambda})}e^{j2\pi\Delta fi(i+\lambda)}\sum_{m = 1}^{L}\!x(i\Delta t - \tau_m)h_m(i\Delta t)\sum_{m = 1}^{L}\!x((i+\lambda)\Delta t - \tau_m)h_m((i+\lambda)\Delta t)\Bigg]^2
\end{align}
\hrulefill
\vspace*{4pt}
\end{figure*}

\subsection{The Gaussianity Test and Decision Mechanism}
\label{TTfG}
It is known from \cite{Lii} that asymptotic Gaussianity of the summable cumulants leads to
\begin{equation}\label{eqn:13}
    \sqrt{N}(\hat{c}_{4y} - c_{4y}) \overset{dist.}{\underset{N\to\infty}{\sim}} D^r(0,\Sigma_c)
\end{equation} 
\noindent where $D^r$ denotes real Gaussian distribution, $c_{4y}$ is the 4th-order cumulant vector which holds the theoretical cumulant values $c_{4y} \triangleq \lim_{N\rightarrow\infty} E\{\hat{c}_{4y}\}$, and $\Sigma_c$ is the asymptotic covariance matrix of $c_{4y}$ which has the close form of
\begin{equation}\label{eqn:14}
    \Sigma_c \triangleq \lim_{N \to \infty} N\mathrm{E}\{(\hat{c}_{4y} - c_{4y})(\hat{c}_{4y} - c_{4y})'\}.
\end{equation}
\noindent where $'$ denotes the transpose operation. Therefore, the time-domain Gaussianity test can be formulated as a binary hypothesis problem of
\begin{align}\label{eqn:17}
                               & \mathcal{H}_0: \hat{c}_{4y} \backsim D^r(0, N^{-1}\Sigma_c) \nonumber\\
                               & vs.\nonumber\\
                               & \mathcal{H}_1: \hat{c}_{4y} \backsim D^r(c_{4y}, N^{-1}\Sigma_c), \backsim c_{4y} \neq 0.
\end{align}

To be able to solve the problem, in this study, the $d_{G,4}$ test which was proposed in \cite{Giannakis} is adopted to the peculiar case of OFDM signal identification under multipath wireless channels. Therefore the binary hypothesis problem which is given in eqn. \eqref{eqn:17} can be addressed by a chi-square test called as $d_{G,4}$ which is defined as 
\begin{equation}\label{eqn:22}
d_{G,4} \triangleq N\hat{c}_{4y}\hat{\Sigma}_c^{-1}\hat{c}_{4y},
\end{equation}
\noindent and from \cite{Giannakis} it can be seen that
\begin{equation}\label{eqn:23}
    d_{G,4} = N\hat{c}^{'}_{4y}\hat{\Sigma}_c^{-1}\hat{c}_{4y} \overset{dist.}{\underset{N\to\infty}{\backsim}} \chi^2_{N_c}.
\end{equation}
\noindent The pseudo-inverse of $\hat{\Sigma}^{-1}$ should replace inverse in the equations below if $\hat{\Sigma}$ rank-deficient. Therefore, for an $\alpha$ level significance the test in eqn. \eqref{eqn:17} becomes a chi-squared test which can be defined by
\begin{equation}\label{eqn:24}
    d_{G,4} \overset{\mathcal{H}_1}{\underset{\mathcal{H}_0}{\gtrless}} t_\mathit{G} = \chi^2_{N_c}(\alpha).
\end{equation}
\noindent $t_\mathit{G}$ is $\chi^2$ disribution table value with degree of freedom of $\alpha$. The probability of the false alarm is given by
\begin{equation}\label{eqn:25}
    P_F \triangleq \alpha \leq P[d_{G,4} \geq\chi^2_{N_c}|\mathcal{H}_0]
\end{equation}
Under $\mathcal{H}_1$, the distribution of $d_{G,4}$ is estimated from eqn. \eqref{eqn:17} as
\begin{equation}\label{eqn:26}
    d_{G,4} \sim D^r[N\hat{c}^{'}_{4y}\hat{\Sigma}_c^{-1}\hat{c}_{4y}, 4N\hat{c}^{'}_{4y}\hat{\Sigma}_c^{-1}\hat{c}_{4y}]
\end{equation}
\noindent and the probability of detection is given by
\begin{equation}\label{eqn:27}
    P_D \triangleq \alpha \leq P[d_{G,4} \geq t_\mathit{G}|\mathcal{H}_1].
\end{equation}

The fourth order cumulant and covariance matrix estimation processes derived herein is utilized by the introduced asymptotic chi-squared CFAR test. Thus the effects of fading channels which can lead to stronger non-Gaussian components are minimized by the involvement of covariances of the cumulants. Selection of the threshold is conducted by employing the $\chi^2$ test tables based on the degree of freedom of the distribution. On the other hand, fourth order cumulant covariance matrix estimate computation is a long and complex procedure which requires involvement of products of the cross moment terms. Moreover, computation of the $\Sigma_c$ can become impractical when $\hat{c}_{4y}$ vector becomes very large. However, peculiar to the OFDM signals, under the conditions defined in Section \ref{FOCOS}, covariance matrix estimate calculation becomes a simpler process. The general form of covariance estimates of the cumulants for OFDM signals is derived based on the computations in \cite{Brillinger1} (eqn. 2.3.8) and \cite{Giannakis} and reduced to three equations which are consist of several moments of the received signal. The computation process can be found in the Appendix. Furthermore, the test is originally applied to real processes and in this paper the same approach is maintained because the distributional characteristics of the complex signals are also retained by the imaginary and real parts. Besides, operating in real or imaginary domain reduces the computation time for the test output when compared to complex domain. Therefore, the analysis for wireless OFDM signals can be given over either of these domains and the analysis can be extended to complex domain \cite{Akmouche}. The proposed OFDM signal identification method can be executed as follows:
\begin{itemize}
\item calculate the estimates of the fourth order cumulants of the received signal based on given degree of freedom, i.e., $M$, using equation (\ref{eqn:39443}),
\item calculate the $\Sigma_c^{-1}$ of the fourth order cumulants based on the equations given in Appendix,
\item calculate the chi-squared test result from the equation (\ref{eqn:22}), using the estimates computed in the previous two steps, and 
\item compare the test result with the tabular value of the test with the given degree of freedom and make the decision based on equation (\ref{eqn:24}).
\end{itemize} 

The calculation of the estimates of the cumulants at the first step of the proposed method is achieved by a set of multiplication, shifting, and summation operations for a limited number of lags. Moreover in the Appendix, the estimation of the covariance matrix in the second step of the proposed method is reduced down to calculation of a set of high-order moments which does not include any shifting operations. These mathematical operations can be implemented over microprocessors or micro-controllers of the measurement devices by the utilization of smart mathematical algorithms which avoid calculation duplications. Therefore, the implementation of the proposed method can be greatly simplified. Furthermore, the construction of the decision process at the last two steps consist of very straightforward multiplication and comparison operations which do not introduce an additional complexity to implementation of the proposed method over the wireless communications systems.   
\section{Measurement Setup}
\label{MS}

\begin{figure*}[t]
\centering{\subfigure[Measurement setup: transmitter, receiver, laptop computer, router and their connections.]{\includegraphics[width=.35\textwidth]{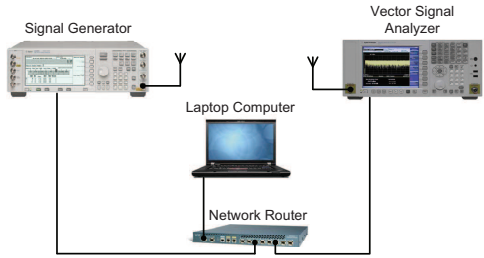}
\label{fig7}}
\hfil
\subfigure[A general overview of the measurement environment.]{\includegraphics[width=.35\textwidth]{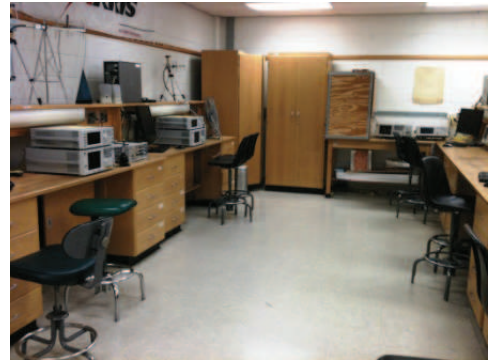}
\label{fig8}}}
\caption{The measurement setup and environment.}
\label{fig_msetup}
\end{figure*}

\begin{table}
\centering
\begin{tabular}{lr}
\hline
Parameter & Value\\
\hline
Carrier frequency & 2140 MHz\\
$T_x/R_x$ distance & $18.24$ feet\\
Modulation type & SC-PSK, SC-FSK, SC-QAM, OFDM-QAM\\
Modulation order & 2,4,8,16,32,64,128,256\\
Number of symbols & 64,128,256,512,1024,2048\\
SNR level & -4 to 15 dB\\
Symbol rate & 250 kHz\\
Filter type & Root raised cosine, $\alpha = 0,3$\\
Frequency Offset ($\Delta f$) & $10$ Hz (max.)\\
\hline
\end{tabular}
\caption{Measurement setup: the values of the used parameters}
\label{tab:1}
\end{table}
A measurement setup is developed to investigate the identification performance of the proposed method. The measurement system consists of an Agilent ESG E4438C signal generator ($T_x$), an Agilent E4440A PSA series spectrum analyzer ($R_x$) and accompanying vector signal analyzer (VSA) software to record the signals, a tuple of omni-directional antennas to cover the wireless communications bands of interest, a laptop computer, cables, and connectors. All the measurement process is controlled by the laptop computer. The signal generator and the spectrum analyzer are connected to the network router and the laptop computer commands both devices through the transmission control protocol and internet protocol. The signal transmission and the reception starts and ends with the commands that are sent by the laptop computer. The setup is depicted in Fig.~\ref{fig7}. Moreover, the signal generator supports the generation and transmission of waveforms that are designed separate from the generator. Therefore, a signal generation software is developed in MATLAB to be able to obtain the full control of the data generation and transmission procedures.

The developed software generates the SC-PSK, SC-FSK, SC-QAM, and OFDM modulated signals with various modulation orders based on the signal model in Section \ref{SM}. The OFDM signals employ QAM as the underlaying modulation technique over $64$ active carriers with $250$ kHz spacing. $T_G$ is also selected as $1/4T_D$. The signal generation software generates the designed waveforms at the laptop and uses vendor provided library functions to access the signal generator. As the software accesses the generator, it uploads generated waveforms to the device and commands the signal generator to transmit the predefined number of symbols of the waveform at the defined frequency band, with the predefined signal power. Moreover, signal parameters such as modulation type and order can also be adjusted by the software. Therefore the parametric analysis which is based on the measurement parameters such as SNR, modulation type, data length, and modulation order can be done methodologically. On the other hand, the VSA software provided by the vendor is employed for signal acquisition and recording purposes. SNR of the received signal is also determined by this software. Therefore, all the signal transmission loop is controlled and completed on the laptop computer while the transmitter and the receiver is allocated to different locations in the laboratory. The list of the all parameters used in the measurement process is given in TABLE~\ref{tab:1}. The measurements are repeated for $200$ times for each changing parameter by the aid of signal generation software.

The measurement environment also has an important effect over the performance of the proposed method. The $T_x/R_x$ devices are fixed to certain allocations in the measurement laboratory which has scattering surfaces such as metal panels, other measurement and calibration equipment, tables, chairs etc., as shown in Fig.~\ref{fig8}. Therefore, the measurement environment is a highly reflective indoor environment. It should be noted that the performance of any signal identification method will be affected under these conditions, because the wireless channel is defined by the heavily reflective nature of the indoor environment. These circumstances can be assumed to be the bottom line performance conditions and it can also be assumed that the identification performance of the proposed method will increase as the system is deployed outdoors with relatively less reflections.

\section{Measurement Results}
\label{MR}
\begin{figure}[t]
\centering
\includegraphics[width=\columnwidth]{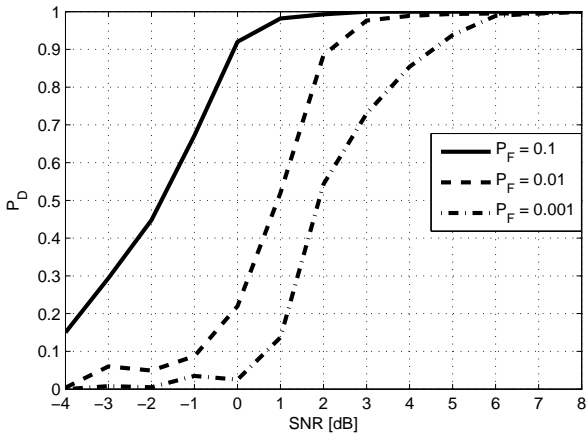}
\caption{Probability of detection for SC-FSK v.s. OFDM: $N = 1024$, modulation order = $32$, $M = 1.5T_S$.}
\label{fig1}
\end{figure}

\begin{figure}[t]
\centering
\includegraphics[width=\columnwidth]{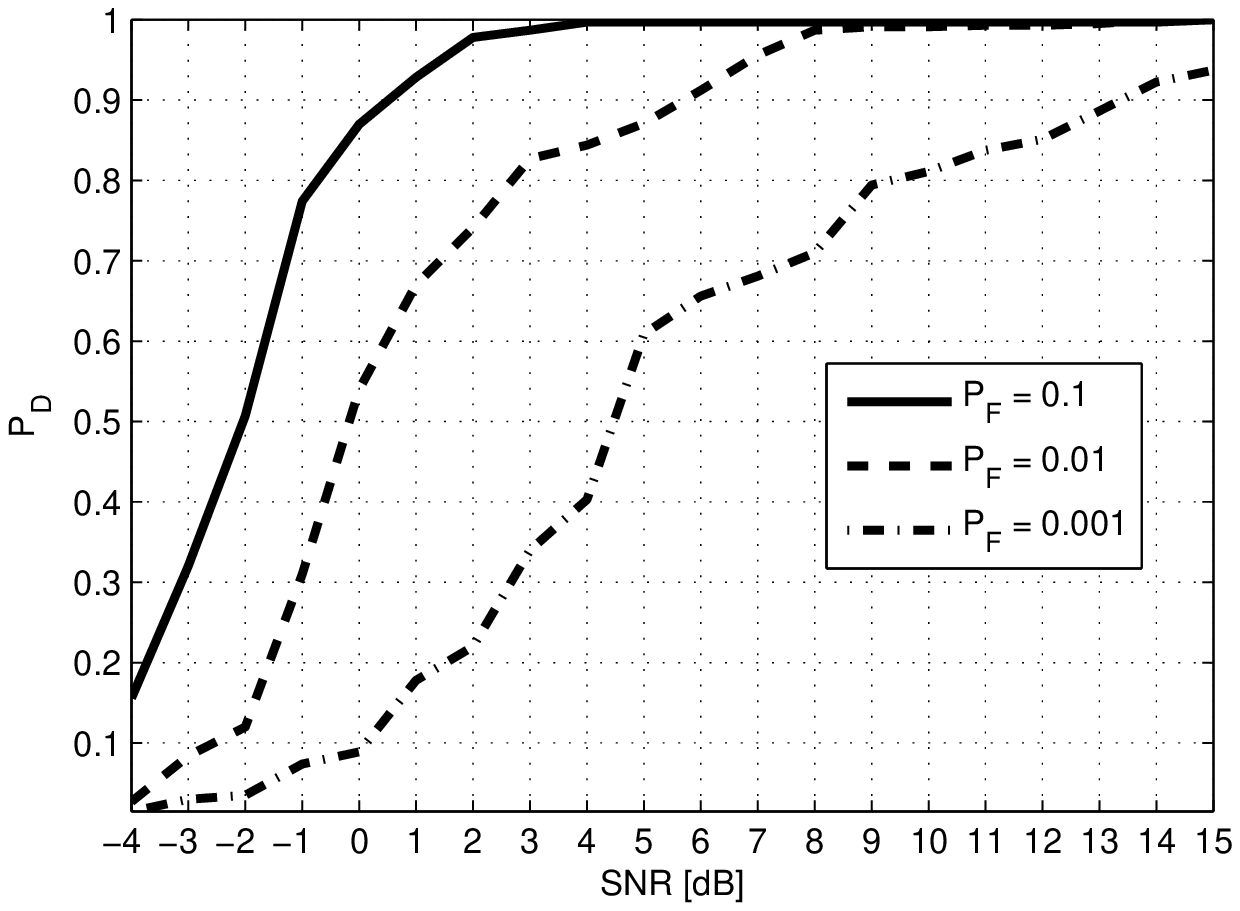}
\caption{Probability of detection for SC-PSK v.s. OFDM: $N = 1024$, modulation order = $32$, $M = 1.5T_S$.}
\label{fig2}
\end{figure}

\begin{figure}[t]
\centering
\includegraphics[width=\columnwidth]{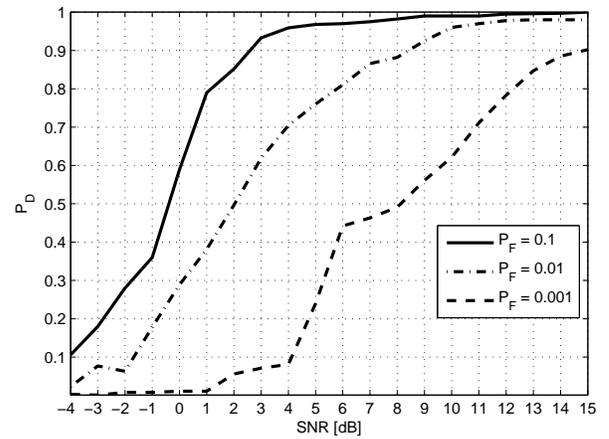}
\caption{Probability of detection for SC-QAM v.s. OFDM: $N = 1024$, modulation order = $32$, $M = 1.5T_S$.}
\label{fig3}
\end{figure}

A CFAR detector is implemented based on the proposed method and the equations \eqref{eqn:22}-\eqref{eqn:27} in Section \ref{TTfG} for $P_F$ of $0.001$, $0.01$, $0.1$ and $200$ measurements. On the other hand, a frequency shift of $\Delta f = 10$ Hz is added to the signal generation process to imitate the Doppler shift and other impairments that cause the shift in the frequency. In Fig.~\ref{fig1}, \ref{fig2}, and \ref{fig3} the $P_D$ results for OFDM vs. SC signals are given for $1024$ symbols, modulation order of $32$, $M = 1.5T_S$, and various SNR levels. The results indicate that best separation is achieved against SC-FSK, SC-PSK and SC-QAM consecutively. Especially for $P_F$ of $0.1$, and $0.01$, a certain confidence level is achieved at $2$ dB of SNR for SC-FSK signals, at $6$ dB for SC-PSK, and at $9$ dB for SC-QAM.

The detection performance of the proposed method falls sharply in a range of couple of dBs at the end of positive SNR region for all signals and identification becomes impossible when the SNR is negative. The measurement results indicate that the noise dominates the test statistics in this region, therefore identification performance deteriorates. SC-QAM follows other modulation types in terms of performance in all measurements because the underlying modulation that is employed by the OFDM signal is also QAM. For the modulation order of $32$, $1024$ symbols, and $P_F = 0.1$, detection performance above $0.9$ is achieved at $0$ dB of SNR for OFDM vs. SC-FSK, at $2$ dB for OFDM vs. SC-PSK, and at $3$ dB for OFDM vs. SC-QAM.

\begin{figure}[t]
\centering
\includegraphics[width=\columnwidth]{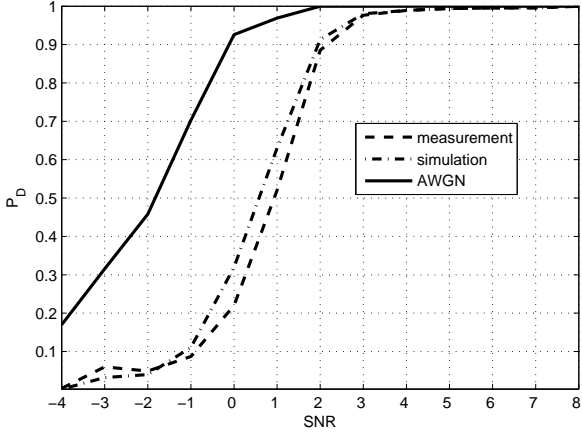}
\caption{$P_D$ performance comparison for SC-FSK v.s. OFDM: AWGN channel, simulations, measurements. $(P_F = 0.01)$, $N = 1024$, modulation order  = $32$, $M = 1.5T_S$}
\label{fig11}
\end{figure}

\begin{figure}[t]
\centering
\includegraphics[width=\columnwidth]{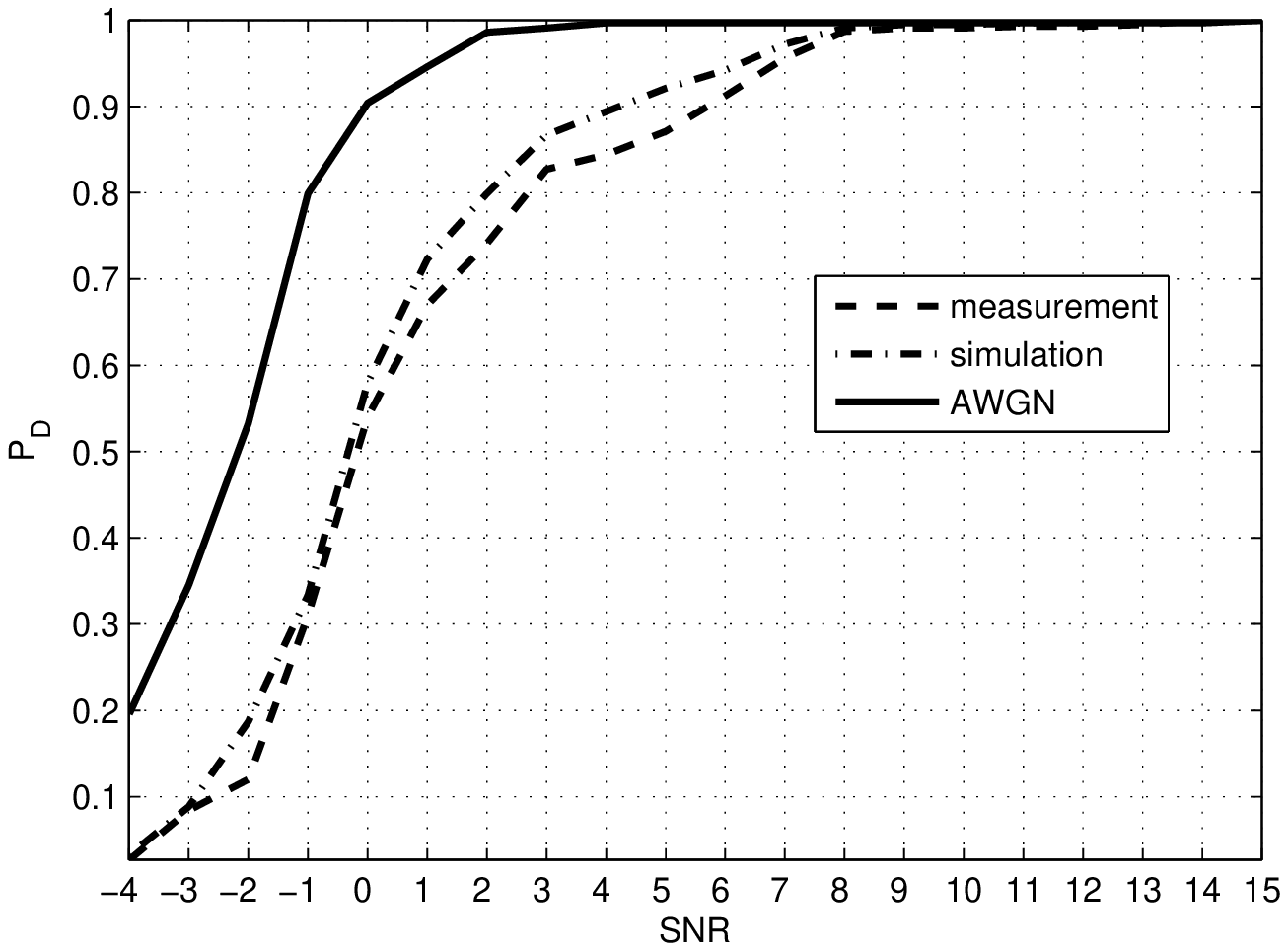}
\caption{$P_D$ performance comparison for SC-PSK v.s. OFDM: AWGN channel, simulations, measurements. $(P_F = 0.01)$, $N = 1024$, modulation order  = $32$, $M = 1.5T_S$}
\label{fig12}
\end{figure}

\begin{figure}[t]
\centering
\includegraphics[width=\columnwidth]{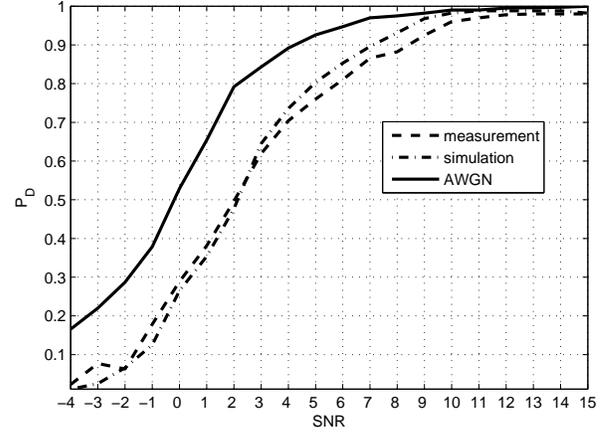}
\caption{$P_D$ performance comparison for SC-QAM v.s. OFDM: AWGN channel, simulations, measurements. $(P_F = 0.01)$, $N = 1024$, modulation order  = $32$, $M = 1.5T_S$}
\label{fig13}
\end{figure}

The measurement results are compared with the simulation results which are based on \cite{Gorcin} for each signal type and $P_F = 0.01$ in Fig.~\ref{fig11}, \ref{fig12}, and \ref{fig13}. Except for a set of negative SNR values, in general the comparison plots indicate that simulation results are slightly better and in some cases -especially in the initial positive SNR region, an increase of $0.1$ at the $P_D$ values can be observed for all signals. This is the indication of the significant effect of measurement environment depicted in Section \ref{MS} over the wireless channel and consequently on the performance of the proposed method in contrary to the four taps simulated channel implemented in \cite{Gorcin}. On the other hand, transmitter and the receiver are directly connected with a cable and the measurement process is synchronized by the available software libraries in both of the devices. These conditions imply only the effect of AWGN is introduced to the received signal during the transmission, thus the plots that are given in the same figures with the label of AWGN can be considered as the upper bound performance limits for the proposed signal identification method. These measurement indicate $P_D$ above $0.9$ for $0$ dB of SNR in case of FSK vs. OFDM, $1$ dB in case of PSK vs. OFDM, and $5$ dB in case of QAM vs. OFDM.

The $P_D$ performance of the proposed method is also investigated by employing QPSK modulated SC wideband code division multiple access (WCDMA) and OFDM based LTE signals. The transmission bandwidth for WCDMA is 5 MHz and LTE system is also configured to 5 MHz transmission bandwidth with 15 KHz spaced carriers and fast Fourier transform size of 512. The signals are generated from the wireless standards based libraries available in the Agilent ESG E4438C signal generator. The rest of the measurement parameters are kept same as given in TABLE~\ref{tab:1}. The measurement results show similarity with that of generic SC-PSK measurements however, $P_D$ leads to certain identification at higher SNR values when compared to the generic case.  

The effect of the change of modulation order on the test statistics is investigated in Fig.~\ref{fig4} when the SNR is $7$ dB, $P_F = 0.1$, and $1024$ symbols are recorded. While SC-FSK separation barely depends on the modulation order, as the order increase above $8$, the proposed method performs better for SC-PSK steadily, and $P_D$ reaches to the levels above $0.95$. In case of SC-QAM signals, the best performance is achieved as the modulation order reaches to $32$ with $P_D > 0.96$. The measurement results indicate no significant effect of modulation order over identification performance.

\begin{figure}[t]
\centering
\includegraphics[width=\columnwidth]{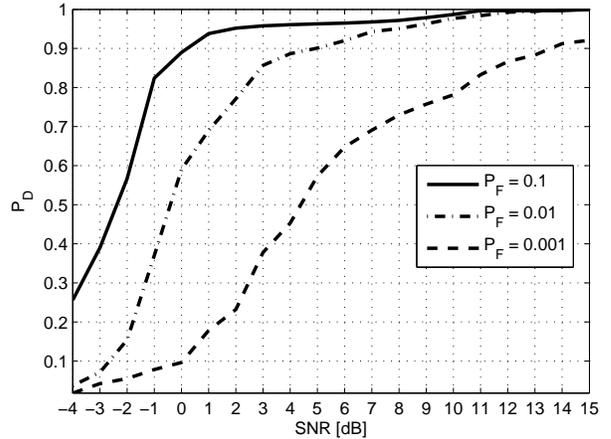}
\caption{Probability of detection for WCDMA v.s. LTE: $M = 1.5T_S$.}
\label{fig10}
\end{figure}

Data length or number of symbols used is also another important parameter which affects the performance of the statistical tests. Fig.~\ref{fig5} displays the detection performance of the proposed method with the changing number of symbols for $7$ dB SNR, $P_F = 0.1$, and modulation order of $32$ for each signal type. Below $128$ symbols the proposed method performs poorly due to the problems discussed in Section~\ref{FOCoOS}.  While the test performs well for FSK signals over $256$ symbols with $P_D > 0.95$, for all SC signals the test performance is optimized as the number of symbols reaches $512$. The degradation in the test performance after $1024$ symbols is due to central limit theorem.

\begin{figure}[t]
\centering
\includegraphics[width=\columnwidth]{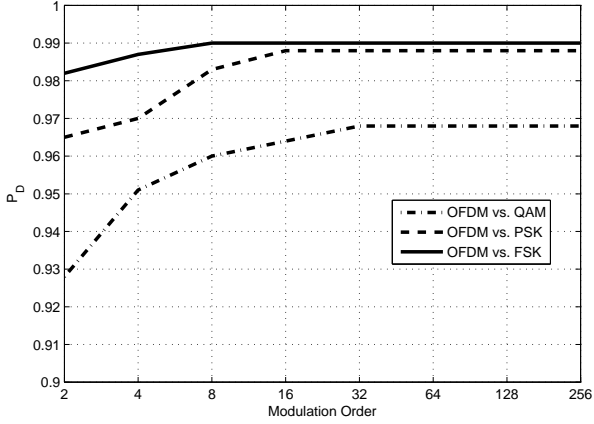}
\caption{Effect of modulation order on the identification: $(P_F = 0.1)$, SNR = $7$ dB, $N = 1024$, $M = 1.5T_S$.}
\label{fig4}
\end{figure}

The lag distance which also determine the degree of freedom of the test has an important effect on the identification performance of the proposed method. More information about the distribution of the received signal is introduced to the decision process as the lag distance increases. However, the computation time required to calculate the estimates of the cumulants will also increase as the lag distance increases. The detection improvements for increased lag distance in terms of symbol duration is given in Fig.~\ref{fig9} for $N = 1024$, SNR = $7$ dB, modulation order  of $32$ and $P_F = 0.01$. The oversampling rate is double the Nyquist rate, therefore, for instance, $M = 0.5T_S$ corresponds to $2$ samples. A significant gain of $0.4$ in the detection performance can be noticed in case of QAM signals as the lag distance increase from $0.5$ to $2.5$, however computation time requirements also increase significantly. For other signal types the $P_D$ performances are steady \textit{i.e.}, above $0.9$.

\begin{figure}[t]
\centering
\includegraphics[width=\columnwidth]{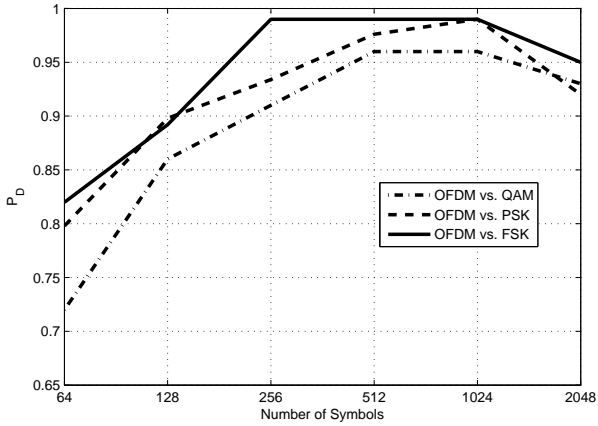}
\caption{Effect of number of symbols on the identification: $(P_F = 0.1)$, SNR = $7$ dB, modulation order = $32$. $M = 1.5T_S$.}
\label{fig5}
\end{figure}

Finally the identification performance of the proposed method is compared with three of the non-coherent statistical methods and a coherent method that are available in the literature. One of these methods, which is introduced in \cite{WeiDai} depends on the fourth and the sixth orders of moments of the OFDM signals and introduces an MMSE algorithm based on these moment functions (MMSE-M). Secondly in \cite{Ulovec}, a method based on the combination of fourth order moments and the amplitude estimation is introduced (AE-M) and third, sums of fourth order cumulants are used for OFDM v.s. SC classification in \cite{Shi} (SUM-C). The coherent method employs a two stage identification procedure to distinguish SLCD, BT-SLCD, and OFDM signals from each other based on the unique second order cyclostationary features of these signals. First the SLCD signals are distinguished from the BT-SLCD/OFDM pair, then these two signals are identified in the second stage \cite{QZhang} (BT-SLCD), however the first stage of the signal classification algorithm is investigated for the comparison purposes assuming that the there are two possible signal types. The performance comparison with these methods can be found in Fig.~\ref{fig6} for the SC-PSK case. Modulation order is selected as $32$, $P_F = 0.01$, and $N = 1204$. 

The proposed method performs better than the non-coherent methods because of the underlying test structure which does not depend on fixed thresholds and utilization of the estimates of the covariances of the cumulants in the identification process. At $1$ dB SNR region, the proposed method outperforms other methods with the $P_D$ margin of $0.2$. For the $5$ dB of SNR, the margin is $0.15$ with SUM-C, $0.08$ with AE-M, and $0.05$ with MMSE-M. As the SNR reaches to $10$ dB, the tests converge to closer $P_D$ rates. On the other hand, while the coherent BT-SLCD algorithm outperforms the proposed method on the low-SNR region, the identification performance of the proposed method converges to the BT-SLCD around the $8$ dB SNR region. These results are expected when it is considered that in general coherent techniques have better performance than the statistical test, trading a certain level of increase in the algorithm complexity and availability of certain information to the receiver beforehand with improvements in terms identification performance. The performance comparison for SC-FSK and SC-QAM signals are not included, but similar results are observed.

\begin{figure}[t]
\centering
\includegraphics[width=\columnwidth]{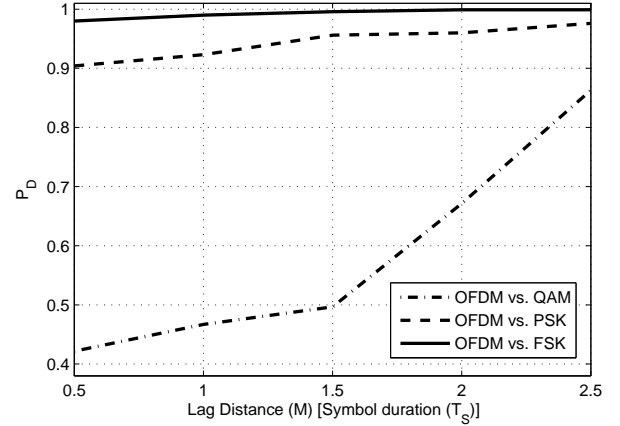}
\caption{Effect of lag distance (degree of freedom) on the identification: $(P_F = 0.01)$, $N = 1024$, SNR = $7$ dB, modulation order  = $32$.}
\label{fig9}
\end{figure}

\begin{figure}[t]
\centering
\includegraphics[width=\columnwidth]{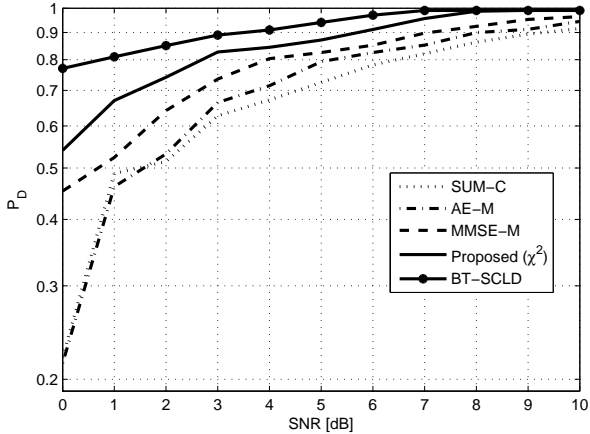}
\caption{Performance comparison with other algorithms: SC-PSK v.s. OFDM, $(P_F = 0.01)$, $N = 1024$, modulation order  = $32$, $M = 1.5T_S$.}
\label{fig6}
\end{figure} 

\section{Conclusion}
\label{C}
Signal identification is a vital component of modern wireless communications systems. In this paper first, fourth order cumulant estimates of wireless OFDM signals are given. Secondly, it is shown that wireless impairments have significant impacts on the distribution characteristics of wireless signals. Then, an OFDM signal identification method which employs estimates of fourth order cumulants and their covariances for limited number of lags is proposed. The cumulant and the covariance matrix calculations are greatly simplified particularly for OFDM signals. The method utilizes a modified $\chi^2$ test at the decision process and symbol duration information is used to determine the number of cumulant legs or the degree of freedom of the underlying Gaussianity test. In practice, degree of freedom which is $M > 10$ can satisfy certain confidence levels, thus the requirement for the symbol duration can also be waived. A measurement setup is developed to validate the performance of the proposed method realistically. The identification performance of the method is provided utilizing parameters such as modulation order and type, data length, and degree of freedom of the $\chi^2$ test. A CFAR detector is used in the performance analysis. The method performs better than other statistical identification methods under the wireless multipath fading channel conditions and it is robust against the changing modulation order and data length. However the identification performance is affected under low-SNR, because of the time-domain Gaussianity of the OFDM signals. $P_D$ of more than $0.9$ is achieved for all signal types when $P_F = 0.1$.

\appendix 
The asymptotic covariances of sample cumulants can be estimated based on the straightforward asymptotic covariance expression for the cumulant orders up to $k = 3$ \cite{Giannakis2}. However, the covariance estimation becomes a complex procedure for $k > 3$ due to the combinations of products of the second and fourth order moments. The estimate of the 4-th order cumulants can be written in terms of moments as \cite{Giannakis}
\begin{align}\label{eqn:28}
  \hat{c}_{4y}(i_1,i_2,i_3) = &\ \hat{m}_{4y}(i_1,i_2,i_3) -  \hat{m}_{2y}(i_1)\hat{m}_{2y}(i_2 - i_3)\nonumber\\
                              & - \hat{m}_{2y}(i_2)\hat{m}_{2y}(i_3 - i_1) - \hat{m}_{2y}(i_3)\hat{m}_{2y}(i_1 - i_2)\tag{A.1}
\end{align}
\noindent where
\begin{equation}\label{eqn:29}
     \hat{m}_{ky}(i_k) \triangleq \frac{1}{N}\sum_{t=0}^{T-1}y(t)y(t+i_1)y(t+i_2)\dots y(t+i_{k-1}),\tag{A.2}
\end{equation}
\noindent and $i_k \triangleq (i_1, i_2,\dots,i_{k-1})$. The computation of each entry $cov\{\hat{c}_{4y}(i_1,i_2,i_3),\hat{c}_{4y}(j_1,j_2,j_3)\}$ of the $\Sigma_c$ becomes a computationally consuming process under $I^N_4$. However the computation of the diagonal entries defined by $I^M_4$ is sufficient for OFDM signals. If $\lambda = i_1 = i_2$, $\beta = j_1 = j_2$, $i_3 = j_3 = 0$ and $0 \leq \beta \leq M$, the open for of covariance of $\hat{c}_{4y}$ can be given in three parts and the first part is consists of covariances of two fourth order moments as 
\begin{align}
c_1 = cov\{\hat{m}_{4y}(\lambda,\lambda,0),\hat{m}_{4y}&(\beta,\beta,0)\}\nonumber\tag{A.3a}\label{eqn:32},
\end{align}
\noindent and the second part will be consist of various combinations of covariances of fourth order and second order moments. The open expression of these terms are given by
\begin{align}
c_2 =  cov\{\hat{m}_{4y}(\lambda,\lambda,0),\hat{m}_{2y}&(\beta)\hat{m}_{2y}(\beta)\}\tag{A.3b1}\\\label{eqn:33}
+\ cov\{\hat{m}_{4y}(\lambda,\lambda,0),\hat{m}_{2y}&(\beta)\hat{m}_{2y}(-\beta)\}\tag{A.3b2}\\\label{eqn:34}
+\ cov\{\hat{m}_{4y}(\lambda,\lambda,0),\hat{m}_{2y}&(0)\hat{m}_{2y}(0)\}\tag{A.3b3}\nonumber\\\label{eqn:35}
+\ cov\{\hat{m}_{4y}(\beta,\beta,0),\hat{m}_{2y}&(\lambda)\hat{m}_{2y}(\lambda)\}\tag{A.3b4}\\\label{eqn:36}
+\ cov\{\hat{m}_{4y}(\beta,\beta,0),\hat{m}_{2y}&(\lambda)\hat{m}_{2y}(-\lambda)\}\tag{A.3b5}\\\label{eqn:37}
+\ cov\{\hat{m}_{4y}(\beta,\beta,0),\hat{m}_{2y}&(0)\hat{m}_{2y}(0)\}\tag{A.3b6},
\end{align}
\noindent and finally the third term will be constituted by the covariances of all possible second order moments of the estimated cumulants as
\begin{align}
c_3 =  cov\{\hat{m}_{2y}(\lambda)\hat{m}_{2y}(\lambda),\hat{m}&_{2y}(\beta)\hat{m}_{2y}(\beta)\}\tag{A.3c1}\\\label{eqn:39}
+\ cov\{\hat{m}_{2y}(\lambda)\hat{m}_{2y}(\lambda),\hat{m}&_{2y}(\beta)\hat{m}_{2y}(-\beta)\}\tag{A.3c2}\\\label{eqn:40}
+\ cov\{\hat{m}_{2y}(\lambda)\hat{m}_{2y}(\lambda),\hat{m}&_{2y}(0)\hat{m}_{2y}(0)\}\tag{A.3c3}\\\label{eqn:41}
+\ cov\{\hat{m}_{2y}(\lambda)\hat{m}_{2y}(-\lambda),\ & \hat{m}_{2y}(\beta)\hat{m}_{2y}(\beta)\}\tag{A.3c4}\\\label{eqn:42}
+\ cov\{\hat{m}_{2y}(\lambda)\hat{m}_{2y}(-\lambda),\ & \hat{m}_{2y}(\beta)\hat{m}_{2y}(-\beta)\}\tag{A.3c5}\\\label{eqn:43}
+\ cov\{\hat{m}_{2y}(\lambda)\hat{m}_{2y}(-\lambda),\ & \hat{m}_{2y}(0)\hat{m}_{2y}(0)\}\tag{A.3c6}\\\label{eqn:44}
+\ cov\{\hat{m}_{2y}(0)\hat{m}_{2y}(0),\ \hat{m}&_{2y}(\beta)\hat{m}_{2y}(\beta)\}\tag{A.3c7}\\\label{eqn:45}
+\ cov\{\hat{m}_{2y}(0)\hat{m}_{2y}(0),\ \hat{m}&_{2y}(\beta)\hat{m}_{2y}(-\beta)\}\tag{A.3c8}\\\label{eqn:46}
+\ cov\{\hat{m}_{2y}(0)\hat{m}_{2y}(0),\ \hat{m}&_{2y}(0)\hat{m}_{2y}(0)\}.\tag{A.3c9}
\end{align}
Therefore the covariance of $\hat{c}_{4y}$ is given by
\begin{equation}\label{eqn:68}
cov\{\hat{c}_{4y}(\lambda,\lambda,0),\hat{c}_{4y}(\beta,\beta,0)\} = c_1 +c_2 + c_3.\tag{A.3d}
\end{equation}

The derivation of these three groups can be based on the computable sample approximations of the asymptotic covariances established in \cite{Lehmann,Dandawate} for the mixing discrete-time stationary processes such as defined in eqn. \eqref{eqn:10}. The covariance of the moments is defined as
\begin{align}\label{eqn:47}
 \lim_{N \to \infty}Ncov\{\hat{m}_{ky}(i_k),\hat{m}_{ky}(j_l)\} = &\sum_{-\infty}^{\infty} cov\{y(0)y(i_1)\dots y(i_{k-1}),\nonumber\\
y(\tau)y(\tau+j_1)\dots y(\tau+j_{l-1})\}.\tag{A.4}
\end{align}
\noindent Following $cov\{ab\} = E\{ab\} - E\{a\}E\{b\}$ based on eqn. \eqref{eqn:47} and replacing the moments with their estimates as defined in eqn. \eqref{eqn:29}, $(A.3a)$ can be written as
 \begin{align}\label{eqn:48}
c_1 = cov\{\hat{m}_{4y}(\lambda,\lambda,0),\hat{m}_{4y}&(\beta,\beta,0)\}\nonumber\\
\approx \frac{1}{N}\sum_{\tau = -K_N}^{K_N}\big[&\hat{m}_{8y}(\lambda,\lambda,0,\tau,\tau+\beta,\tau+\beta,\tau)\nonumber\\
\ -&\ \hat{m}_{4y}(\lambda,\lambda,0)\hat{m}_{4y}(\beta,\beta,0)\big]\tag{A.5}
\end{align}
Limits of the sum in the eqn. \eqref{eqn:48} should satisfy $K_N \to \infty$ as $N \to \infty$ and $K_N/N \to 0$ as $N \to \infty$. In practice, for OFDM systems $K_N$ can be selected as $K_N > \tau$ where $\hat{c}_{2x_{m}}(\tau) \approx 0$ \cite{Giannakis}. The second group of sample estimates include the covariances of the multiplications of the moment terms and theorem $2.3.2$ in \cite{Brillinger1} expresses such terms in the form of
\begin{equation}\label{eqn:49}
cov\{a, bc\} =  cov\{b\}cov\{a,c\} + cov\{c\}cov\{a,b\}\tag{A.6}
\end{equation}
and eqn. \eqref{eqn:32} can be written as
\begin{align}\label{eqn:50}
cov\{\hat{m}_{4y}(\lambda,\lambda,0),\hat{m}_{2y}(\beta)\hat{m}_{2y}(\beta)\}\qquad\qquad&\nonumber\\
= E\{\hat{m}_{2y}(\beta)\}cov\{\hat{m}_{4y}(\lambda,\lambda&,0),\hat{m}_{2y}(\beta)\}\nonumber\\
+ E\{\hat{m}_{2y}(\beta)\}cov\{\hat{m}_{4y}&(\lambda,\lambda,0),\hat{m}_{2y}(\beta)\}\nonumber\\
= 2\Big[E\{\hat{m}_{2y}(\beta)\}cov\{\hat{m}_{4y}(\lambda&,\lambda,0),\hat{m}_{2y}(\beta)\}\Big].\tag{A.7}
\end{align}
\noindent Depending on the estimator given in eqn. \eqref{eqn:29} and covariance definition in eqn. \eqref{eqn:47}, similar to eqn. \eqref{eqn:48}, the covariances in eqn. \eqref{eqn:50} can be expressed in terms of moments as
\begin{align}\label{eqn:51}
\quad cov\{\hat{m}_{4y}(\lambda,\lambda,0),\hat{m}_{2y}(\beta)\hat{m}_{2y}(\beta)\}\nonumber\\
 = 2\big[\hat{m}_{2y}(\beta)\sum_{\tau = -K_N}^{K_N}[\hat{m}_{6y}(\lambda,\lambda,0,\tau&,\tau+\beta) - \hat{m}_{4y}(\lambda,\lambda,0)\hat{m}_{2y}(\beta)]\big].\tag{A.8}
\end{align}
\noindent When the same derivations are applied to the rest of the equations and after some simplifications (\ref{eqn:32} - \ref{eqn:37}) can be written as
\begin{align}\label{eqn:52}
c_2 = \mathrm{2}\Big( \Omega\big(2\hat{m}_{2y}&(\beta) + \hat{m}_{2y}(-\beta)\big) + \Phi\hat{m}_{2y}(\beta) + 2\Psi\hat{m}_{2y}(0)\Big),\tag{A.9}
\end{align}
\noindent where
\begin{equation}\label{eqn:53}
\Omega = \frac{1}{N}\sum_{\tau = -K_N}^{K_N}[\hat{m}_{6y}(\lambda,\lambda,0,\tau,\tau+\beta) - \hat{m}_{4y}(\lambda,\lambda,0)\hat{m}_{2y}(\beta)],\tag{A.10}
\end{equation}
\begin{equation}\label{eqn:54}
\Phi = \frac{1}{N}\sum_{\tau = -K_N}^{K_N}[\hat{m}_{6y}(\lambda,\lambda,0,\tau,\tau-\beta) - \hat{m}_{4y}(\lambda,\lambda,0)\hat{m}_{2y}(-\beta)],\quad \ \tag{A.11}
\end{equation}
and
\begin{equation}\label{eqn:55}
\Psi = \frac{1}{N}\sum_{\tau = -K_N}^{K_N}[\hat{m}_{6y}(\lambda,\lambda,0,\tau,\tau) - \hat{m}_{4y}(\lambda,\lambda,0)\hat{m}_{2y}(0)].\ \tag{A.12}
\end{equation}
\noindent The $I^M_4$ leads to symmetric $\Sigma_c$ therefore for each $\Sigma_c$ entry not all six terms of second groups should be processed, estimation of the first half would be sufficient and the following terms will be replicated as indicated in eqn. \eqref{eqn:52}. Similar to the second group, covariances of multiplications of the estimate moments which consist the third group are in the form of $cov\{ab,cd\}$. According to eqn.(2.3.8) in \cite{Brillinger1} these terms can be expressed as
\begin{equation}\label{eqn:56}
cov\{ab, cd\} = cov\{a,c\}cov\{b,d\} + cov\{a,d\}cov\{b,c\},\tag{A.13}
\end{equation}
\noindent then the eqn. \eqref{eqn:37} is given by
\begin{align}\label{eqn:57}
cov\{\hat{m}_{2y}&(\lambda)\hat{m}_{2y}(\lambda),\hat{m}_{2y}(\beta)\hat{m}_{2y}(\beta)\}\nonumber\\
&= cov\{\hat{m}_{2y}(\lambda),\hat{m}_{2y}(\beta)\}cov\{\hat{m}_{2y}(\lambda),\hat{m}_{2y}(\beta)\}\nonumber\\ 
&\ + cov\{\hat{m}_{2y}(\lambda),\hat{m}_{2y}(\beta)\}cov\{\hat{m}_{2y}(\lambda),\hat{m}_{2y}(\beta)\}.\tag{A.14}
\end{align}
\noindent After applying the conversion defined in the eqn. \eqref{eqn:56} and rearranging all nine terms, (\ref{eqn:37} - \ref{eqn:46}) can be written as
\begin{align}\label{eqn:58}
c_3 =  & 2\Gamma^2 + 2\Gamma\Lambda + 2\Theta^2 + 2\Gamma cov\{\hat{m}_{2y}(\lambda),\hat{m}_{2y}(-\beta)\}\nonumber\\
&\ + \Delta(\Gamma + \Lambda) + 2\Theta cov\{\hat{m}_{2y}(-\lambda),\hat{m}_{2y}(0)\} + 2\Upsilon^2\nonumber\\
&\ + 2\Upsilon cov\{\hat{m}_{2y}(0),\hat{m}_{2y}(-\beta)\} +2\big[cov\{\hat{m}_{2y}(0),\hat{m}_{2y}(0)\}\big]^2,\tag{A.15}
\end{align}
\noindent where 
\begin{equation}\label{eqn:59}
\Gamma = \frac{1}{N}\sum_{\tau = -K_N}^{K_N}[\hat{m}_{4y}(\lambda,\tau,\tau+\beta) - \hat{m}_{2y}(\lambda)\hat{m}_{2y}(\beta)],\ \ \ \tag{A.16}
\end{equation}
\begin{equation}\label{eqn:60}
\Lambda = \frac{1}{N}\sum_{\tau = -K_N}^{K_N}[\hat{m}_{4y}(\lambda,\tau,\tau-\beta) - \hat{m}_{2y}(\lambda)\hat{m}_{2y}(-\beta)],\ \ \ \tag{A.17}
\end{equation}
\begin{equation}\label{eqn:61}
\Theta = \frac{1}{N}\sum_{\tau = -K_N}^{K_N}[\hat{m}_{4y}(\lambda,\tau,\tau) - \hat{m}_{2y}(\lambda)\hat{m}_{2y}(0)],\quad \ \tag{A.18}
\end{equation}
\begin{equation}\label{eqn:62}
\Delta = \frac{1}{N}\sum_{\tau = -K_N}^{K_N}[\hat{m}_{4y}(-\lambda,\tau,\tau-\beta) - \hat{m}_{2y}(-\lambda)\hat{m}_{2y}(-\beta)],\tag{A.19}
\end{equation}
\begin{equation}\label{eqn:63}
\Upsilon = \frac{1}{N}\sum_{\tau = -K_N}^{K_N}[\hat{m}_{4y}(\lambda,\tau,\tau+\beta) - \hat{m}_{2y}(\lambda)\hat{m}_{2y}(\beta)],\quad \ \ \ \tag{A.20}
\end{equation}
\noindent and
\begin{align}\label{eqn:64}
cov\{&\hat{m}_{2y}(-\lambda),\hat{m}_{2y}(\beta)\}\nonumber\\ 
&= \frac{1}{N}\sum_{\tau = -K_N}^{K_N}[\hat{m}_{4y}(-\lambda,\tau,\tau+\beta) - \hat{m}_{2y}(-\lambda)\hat{m}_{2y}(\beta)],\tag{A.21}
\end{align}
\begin{equation}\label{eqn:65}
cov\{\hat{m}_{2y}(-\lambda),\hat{m}_{2y}(0)\} = \frac{1}{N}\sum_{\tau = -K_N}^{K_N}[\hat{m}_{4y}(-\lambda,\tau,\tau) - \hat{m}_{2y}(-\lambda)\hat{m}_{2y}(0)],\tag{A.22}
\end{equation}
\begin{equation}\label{eqn:66}
cov\{\hat{m}_{2y}(0),\hat{m}_{2y}(-\beta)\} = \frac{1}{N}\sum_{\tau = -K_N}^{K_N}[\hat{m}_{4y}(0,\tau,\tau-\beta) - \hat{m}_{2y}(0)\hat{m}_{2y}(-\beta)],\tag{A.23}
\end{equation}
\begin{equation}\label{eqn:67}
cov\{\hat{m}_{2y}(0),\hat{m}_{2y}(0)\} = \frac{1}{N}\sum_{\tau = -K_N}^{K_N}[\hat{m}_{4y}(0,\tau,\tau) - \hat{m}_{2y}(0)\hat{m}_{2y}(0)].\quad\tag{A.24}
\end{equation}
Each entry of $\Sigma_c$ now can be computed as the sum of equations \eqref{eqn:48}, \eqref{eqn:52}, and \eqref{eqn:58} under $I^M_4$.


\begin{thebibliography}{1}
\bibitem{SM.2152} ITU Radiocommunication Sector, ``{Technical Identification of Digital Signals},'' ITU-R SM.1600-1, September 2012.

\bibitem{Swami} A. Swami,  and B.M. Sadler, ``{Hierarchical Digital Modulation Classification Using Cumulants},'' In \emph{IEEE Transactions on Communications,} vol. 48, no. 3, pp. 416-429, Mar 2000.

\bibitem{Grimaldi} D. Grimaldi, S. Rapuano,  and L. De Vito , ``An Automatic Digital Modulation Classifier for Measurement on Telecommunication Networks'', In \emph{IEEE Transactions on Instrumentation and Measurement,} vol. 56, no. 5, pp. 1711-1720, 2007.

\bibitem{Shi} M. Shi, A. Laufer, Y. Bar-Ness, and W. Su, ``{Fourth Order Cumulants in Distinguishing Single Carrier from OFDM Signals},'' In \emph{Proc. IEEE Military Communications Conference (MILCOM 2008),} San Diego, CA, U.S.A., 16-19 Nov. 2008 , pp. 1-6

\bibitem{Liu} W. Liu, J. Wang, S.u  Li, ``Blind Detection and Estimation of OFDM Signals in Cognitive Radio Contexts,'' In \emph{Proc. 2nd International Conference on Signal Processing Systems (ICSPS 2010),} Dalian, China, 5-7 July 2010, pp. 347-351.

\bibitem{Ulovec} K. Ulovec, ``{Recognition of OFDM Modulation Method},'' In \emph{Journal of Radioengineering,} vol. 17, no. 1, pp. 50-55, April 2008.

\bibitem{HWang} H. Wang, B. Li, and Y. Wang, ``{Modulation Identification for OFDM in Multipath Circumstance Based on Support Vector Machine},'' In \emph{Proc. 11th IEEE Singapore International Conference on Communication Systems (ICCS 2008),} Guangzhou, China, 19-21 Nov. 2008, pp. 1349-1353

\bibitem{WeiDai} Wei Dai, Y. Wang, J. Wang ``Joint Power Estimation and Modulation Classification Using Second and Higher Statistics'', In \emph{Proc. IEEE Wireless Communications and Networking Conference (WCNC2002),} vol. 1, Orlando, Florida, U.S.A., 17-21 Mar 2002, pp. 155-158

\bibitem{Wang} Bin Wang, Lindong Ge, ``Blind Identification of OFDM Signal in Rayleigh Channel'', In \emph{Proc. IEEE 5th International Conference on Information, Communications and Signal Processing,} Bangkok, Thailand, 6-9 Dec. 2005, pp. 950-954

\bibitem{Muhlhaus} M. Muhlhaus, Menguc Oner, O. Dobre, and F. Jondral, ``A low complexity modulation classification algorithm for MIMO systems,'' In \emph{IEEE Communications Letters,} vol. 17, no. 10, pp. 1881-1884, 2013.

\bibitem{Leinonen} J. Leinonen,  and M. Juntti, ``{Modulation Classification in Adaptive OFDM Systems},'' In \emph{Proc. IEEE 59th Vehicular Technology Conference (VTC 2004-Spring),} vol.3, Milan, Italy, 17-19 May 2004, pp. 1554-1558

\bibitem{FWang} F. Wang, and B. Li, ``{A New Method for Modulation Classification Based on Bootstrap Technique},'' In \emph{Proc. IEEE International Symposium on Computer Science and Computational Technology (ISCSCT'08),} vol. 2, Shanghai, China, 20-22 Dec. 2008, pp. 11-14

\bibitem{Zhang} J. Zhang, and B. Li, ``{A New Modulation Identification Scheme for OFDM in Multipath Rayleigh Fading Channel},'' In \emph{Proc. IEEE International Symposium on Computer Science and Computational Technology (ISCSCT'08),} vol. 1, Shanghai, China, 20-22 Dec. 2008, pp. 793-796

\bibitem{Haq} K.N. Haq, A. Mansour, S. Nordholm, ``Classification of Digital Modulated Signals Based on Time Frequency Representation,'' In \emph{Proc. 2nd International Conference on Signal Processing Systems (ICSPS 2010),} Dalian, China, 5-7 July 2010, pp. 1-5

\bibitem{Punchihewa} A. Punchihewa, O.A. Dobre, S. Rajan, and  R. Inkol, ``{Cyclostationarity-based Algorithm for Blind Recognition of OFDM and Single Carrier Linear Digital Modulations},'' In \emph{Proc. IEEE 18th International Symposium on Personal, Indoor and Mobile Radio Communications (PIMRC 2007),} Athens, Greece, 3-7 Sept. 2007, pp. 1-5

\bibitem{QZhang} Qiyun Zhang, Octavia A. Dobre, Yahia A. Eldemerdash, S. Rajan, and R. Inkol, ``Second-Order Cyclostationarity of BT-SCLD Signals: Theoretical Developments and Applications to Signal Classification and Blind Parameter Estimation,'' In \emph{IEEE Transactions on Wireless Communications,} vol. 12, no. 4, pp. 1501-1511, 2013.

\bibitem{Ala} Ala Al-Habashna, Octavia A. Dobre, Ramachandran Venkatesan, and Dimitrie C. Popescu, ``Second-order cyclostationarity of mobile WiMAX and LTE OFDM signals and application to spectrum awareness in cognitive radio systems,'' In \emph{IEEE Journal of Selected Topics in Signal Processing,} vol. 6, no. 1, pp. 26-42, 2012.


\bibitem{Bouzegzi} A. Bouzegzi, P. Ciblat, P.Jallon, ``New Algorithms for Blind Recognition of OFDM Based Systems'', In \emph{Elsevier Signal Processing Journal,} vol. 90, no. 3, pp. 900-913, 2010.

\bibitem{Hazza} Alharbi Hazza, Mobien Shoaib, Saleh A. Alshebeili, and Alturki Fahad, ``An overview of feature-based methods for digital modulation classification,'' In \emph{Proc. IEEE 1st International Conference on Communications, Signal Processing, and their Applications (ICCSPA'13),} Sharjah, UAE, 12-14 Feb. 2013, pp. 1-6.

\bibitem{Li} H. Li, Y. Bar-Ness, A.  Abdi, O.S. Somekh, and W. Su, ``{OFDM Modulation Classification and Parameters Extraction},'' In \emph{Proc. IEEE 1st International Conference on Cognitive Radio Oriented Wireless Networks and Communications,} Mykonos Island, Greece, 8-10 June 2006, pp. 1-6

\bibitem{Giannakis} G. B. Giannakis, M. K. Tsatsanis, ``Time-domain Tests for Gaussianity and Time-reversibility'', In \emph{IEEE Transactions on Signal Processing,} vol. 42, pp. 3460-3472, Dec. 1994.

\bibitem{Akmouche} W. Akmouche,``Detection of Multicarrier Modulations Using 4th-order Cumulants'', In \emph{Proc. IEEE Military Communications Conference (MILCOM 1999),} Atlantic City, NJ, U.S.A., 31 Oct 1999 - 03 Nov 1999, pp. 432-436.

\bibitem{Gorcin} A. Gorcin, H. Arslan, ``Identification of OFDM Signals Under Multipath Fading Channels'', In \emph{Proc. IEEE Military Communications Conference (MILCOM 2012),} Orlando, USA, 29 October – 1 November  2012, pp. 1-7.

\bibitem{Gorcin2} A. Gorcin, H. Arslan, ``Signal Identification for Adaptive Spectrum Hyperspace Access in Wireless Communications Systems'', In \emph{IEEE Communications Magazine,} vol. 52, no. 10, pp. 134-145, 2014.

\bibitem{Brillinger2} D. Brillinger and M. Rosenblatt, ``{Computation and Interpretation of \textit{k}th-order Spectra},'' In \emph{Spectral Analysis of Time Series,} B. Harris, Ed. New York: Wiley, 1967,  pp.189-232. 

\bibitem{Mendel} Jerry M. Mendel, ``Tutorial on higher-order statistics (spectra) in signal processing and system theory: theoretical results and some applications'', In \emph{Proceedings of the IEEE,} vol. 79, no. 3, pp. 278-305, 1991.

\bibitem{Brillinger1} D. Brillinger, ``{Time Series, Data Analysis and Theory},'' San Francisco: Holden day, 1981.

\bibitem{Giannakis2} G. B. Giannakis, M. K. Tsatsanis, ``A Unifying Maximum-Likelihood View of Cumulant and Polyspectral Measures for non-Gaussian Signal Classification and Estimation'', In \emph{IEEE Transactions on Information Theory,} vol. 38, pp. 386-406, Mar. 1992.

\bibitem{Martret} C. Le Martret, D. Boiteau,``Modulation Classification by Mean of Different Orders Statistical Moments'', In \emph{Proc. IEEE Military Communications Conference (MILCOM 1997),} vol.3, Monterey, CA , U.S.A. , 02 - 05 Nov 1997, pp. 1387-1391.

\bibitem{Lii} K.S. Lii and M. Rosenblatt, ``{Deconvolution and estimation of transfer function phase and coefficients for non-Gaussian linear processes},'' In \emph{The Annals of Statistics,} vol. 10, no. 4, pp. 1195-1208, 1982. 

\bibitem{Lehmann} E.L. Lehmann, ``{Theory of Point Estimation},'' New York: Wiley, 1983, ch.5.

\bibitem{Dandawate} A. V. Dandawate, G. B. Giannakis, ``Asymptotic Theory of Mixed Time Averages and kth-order Cyclic-moment and Cumulant Statistics'', In \emph{IEEE Transactions on Information Theory,} vol. 41, no.1, pp. 216-232, Jan 1995.
\end{thebibliography}
\end{document}